\documentclass[12pt]{iopart}
\usepackage{iopams,mathptm,times,graphicx}
\makeatletter

\newcounter{theorem}
\@addtoreset{theorem}{section}
\renewcommand\thetheorem{\arabic{section}.\arabic{theorem}}

\newenvironment{theorem}{\par\medskip\noindent\begingroup{\bf Theorem
             \stepcounter{theorem}\thetheorem.}\ \itshape
             \def\@currentlabel{\thetheorem}}{\endgroup\par\medskip}

\newenvironment{remark}{\par\medskip\noindent\begingroup{\bf Remark
             \stepcounter{theorem}\thetheorem.}\
             \def\@currentlabel{\thetheorem}}{\endgroup\par\medskip}

\newenvironment{proof}{\par\noindent{\bf Proof.} }{\proofbox\par\medskip}

\def\proofbox{\hfill{\ensuremath\Box}}

\newdimen\LENB \newdimen\LENW \newdimen\THI
\newdimen\LENWH \newdimen\LENTOT \newcount\N
\def\vbrknlnele#1#2#3{
  \LENB=#1pt \LENW=#2pt \THI=#3pt
  \LENWH=\LENW \divide\LENWH by 2
  \LENTOT=\LENB \advance\LENTOT by \LENW
  \vbox to \LENTOT{
    \vbox to \LENWH{}
    \nointerlineskip
    \vbox to \LENB{\hbox to \THI{\vrule width \THI height \LENB}}
    \nointerlineskip
    \vbox to \LENWH{}
  }}

\def\vbrknln#1{
  \N=#1
  \vcenter{
    \vbox{
      \loop\ifnum\N>0
        \vbox to 4pt{\vbrknlnele{2}{2}{0.1}}
        \nointerlineskip
        \advance\N by -1
      \repeat
  }}}

\def\hbrknlnele#1#2#3{
  \LENB=#1pt \LENW=#2pt \THI=#3pt
  \LENTOT=\LENB \advance\LENTOT by \LENW
  \vcenter{
    \vbox to \THI{
      \hbox to \LENTOT{
        \hfil
        \vrule width \LENB height \THI
        \hfil}
  }}}

\makeatletter
\usepackage{ulem}

\def\journal#1&#2,{\begingroup \let\journal=\dummyjournal
               \it #1\unskip~\bf\ignorespaces #2\rm,\endgroup}
\def\dummyjournal{\errmessage{Reference foul up: nested \journal macros}}

\eqnobysec

\def\eqref#1{(\ref{#1})}

\begin{document}
\title[The two-component reduced Ostrovsky equation]
  {A two-component generalization of the reduced Ostrovsky equation and its integrable semi-discrete analogue}
\author{Bao-Feng Feng$^1$, Ken-ichi Maruno$^2$ and
Yasuhiro Ohta$^{3}$
}
\address{$^1$~School of Mathematical and Statistical Sciences,
The University of Texas Rio Grande Valley,
Edinburg, TX 78539
}

\address{$^2$~Department of Applied Mathematics,
Waseda University, Tokyo 169-8050, Japan
}
\address{$^3$~Department of Mathematics,
Kobe University, Rokko, Kobe 657-8501, Japan
}
\eads{\mailto{baofeng.feng@utrgv.edu}, \mailto{kmaruno@waseda.jp} and
\mailto{ohta@math.kobe-u.ac.jp}}

\begin{abstract}
In the present paper, we  propose a two-component generalization of the reduced Ostrovsky equation, whose differential form can be viewed as the short-wave limit of a two-component Degasperis-Procesi (DP) equation. They are integrable due to the existence of Lax pairs. Moreover, we have shown that two-component reduced Ostrovsky equation can be reduced from an extended BKP hierarchy with negative flow through a pseudo 3-reduction and a hodograph (reciprocal) transform. As a by-product, its bilinear form and $N$-soliton solution in terms of pfaffians are presented. One- and two-soliton solutions are provided and analyzed. In the second part of the paper, we start with a modified BKP hierarchy, which is a B\"acklund transformation of the above extended BKP hierarchy, an integrable semi-discrete analogue of two-component reduced Ostrovsky equation is constructed by defining an appropriate discrete hodograph transform and dependent variable transformations. Especially, the backward difference form of above semi-discrete two-component reduced Ostrovsky equation gives rise to the integrable semi-discretization of the short wave limit of a two-component DP equation.
Their $N$-soliton solutions in terms of pffafians are also provided.
\par
\kern\bigskipamount\noindent
\end{abstract}
\kern-\bigskipamount

{\textbf{Keywords}}: BKP and modified BKP hierarchy; pseudo 3-reduction; hodograph and discrete hodograph transform; two-component reduced Ostrovsky equation; short wave model of two-component Degasperis-Procesi (DP) equation; integrable discretization;

\pacs{02.30.Ik, 05.45.Yv,42.65.Tg, 42.81.Dp}


\section{Introduction}
The partial differential equation
\begin{equation}
\left(u_t+c_0 u_x+ uu_x\right)_x=\gamma u\,,\label{VE}
\end{equation}
is a special case ($\beta=0$) of the Ostrovsky equation
\begin{equation}
\left(u_t+c_0 u_x+uu_x+\beta u_{xxx}\right)_x
=\gamma u\,,\label{ostrovsky}
\end{equation}
which was originally derived as a model for weakly nonlinear surface and internal
waves in a rotating ocean~\cite{Ostrovsky,Stepanyants}.
As pointed in \cite{Parkes}, equation (\ref{VE}) is invariant under the transformation
\begin{equation}
u\rightarrow \mu ^{2}u,\quad \ x\rightarrow \mu x,\quad t\rightarrow \mu
^{-1}t\,, \quad c_0\rightarrow \mu ^{2}c_0\,,
\end{equation}
and under the transformation
\begin{equation}
u\rightarrow -u,\quad t\rightarrow -t,\quad \gamma \rightarrow -\gamma\,.
\end{equation}
Moreover, the linear term $c_0 u_x$ can be eliminated by a Galilean transformation.
Therefore, without loss of generality, we can assume $\gamma=3$ and consider specifically the following equation
 \begin{equation}
\left(u_t+uu_x\right)_x-3u=0\,,\label{vakhnenko}
\end{equation}
which is called  the reduced Ostrovsky equation hereafter.
Several authors derived basically the same model equation from different physical situations
\cite{Hunter,Vakhnenko1,Vakhnenko5}. Particularly, it appears as a model for high-frequency
waves in a relaxing medium \cite{Vakhnenko1,Vakhnenko5}. Therefore the reduced Ostrovsky equation (\ref{vakhnenko}) is sometimes called the Vakhnenko equation ~\cite{Vakhnenko2,Vakhnenko3,Vakhnenko4}, the Ostrovsky-Hunter equation
~\cite{Liu}, or the Ostrovsky-Vakhnenko equation~\cite{Brunelli,Shepelsky}.

Differentiating the reduced Ostrovsky equation (\ref{vakhnenko}) with respect to $x$, we obtain
\begin{equation}
u_{txx}+3u_xu_{xx}+uu_{xxx}-3u_x=0\,,\label{short-DP-eq}
\end{equation}
or in an alternative form
\begin{equation} \label{short-DP-eq2}
m_t+um_x +3m u_x=0\,, \quad m=1-u_{xx}
\end{equation}
Eq. (\ref{short-DP-eq}) or Eq. (\ref{short-DP-eq2})
is known as the short wave limit of the Degasperis-Procesi (DP)
equation~\cite{Hone-Wang,MatsunoPLA}.
The reason lies in the fact that eq. (\ref{short-DP-eq}) can be derived from the DP equation~\cite{DP}
\begin{equation}
U_T+3U_X- U_{TXX}+4UU_X=3U_XU_{XX}+UU_{XXX}\,,\label{DP-eq}
\end{equation}
by taking a short wave limit $\epsilon\to 0$ with
$U=\epsilon^2(u+\epsilon u_1+\cdots)$, $T=\epsilon t$,
$X=\epsilon^{-1}x$.
Based on this connection, Matsuno \cite{MatsunoPLA} constructed $N$-soliton solution of the short wave model of
the DP equation from $N$-soliton solution of the DP equation~\cite{Matsuno-DP1,Matsuno-DP2}. By using the reciprocal link between the reduced Ostrovsky equation and periodic 3-reduction of the B-type or C-type two-dimensional Toda lattice, i.e. the $A_2^{(2)}$ 2D-Toda lattice, multi-soliton solutions to both the reduced Ostrovsky equation (\ref{vakhnenko}) and its differentiation form were constructed by the authors in \cite{FMO-VE}. Furthermore, we constructed an integrable semi-discrete reduced Ostrovsky equation \cite{FMO-VE-discrete} from a modified BKP hierarchy based on Hirota's bilinear approach \cite{HirotaBook}. The integrability and wave-breaking was studied in \cite{Grimshaw2012}. Interestingly, the short wave limit of the DP equation (\ref{short-DP-eq}) also serves as an asymptotic model for propagation of surface waves in deep water under the condition of small-aspect-ratio \cite{Manna2014}. Most recently, the inverse scattering transform (IST) problem for the short wave limit of the DP equation (\ref{short-DP-eq}) was solved by a Riemann-Hilbert approach \cite{Shepelsky}.

In the present paper, we propose and study a two-component generalization of the reduced Ostrovsky equation
\begin{equation}
\left(u_t+uu_x\right)_x
=3u+ c(1-\rho)\,,\label{2compVE1}
\end{equation}
\begin{equation}
\rho_t+(\rho u)_x=0\,,
\label{2compVE2}
\end{equation}
which is shown to be integrable by finding its Lax pair and multi-soliton solution in subsequent sections.
Differentiating Eq. (\ref{2compVE1}) with respect to $x$, we also have
\begin{equation}
m_{t}+m_{x}u+3mu_{x}=c\rho _{x}\, \quad m=1-u_{xx}
\label{2compVE3}
\end{equation}
The system (\ref{2compVE2})--(\ref{2compVE3}) is also integrable, which can be viewed as the short wave limit of a two-component Degasperis-Procesi equation.

The remainder of the present paper is organized as follows. In section 2, we find Lax pairs for two-component reduced Ostrovsky equation and its differential form.  Then in section 3, starting from an extended BKP hierarchy with negative flow and its tau functions, we derive a two-component reduced Ostrovsky equation by a pseudo-3 reduction and an appropriate hodograph transform. Its bilinear form and $N$-soliton solution in parametric form are also given.
In section 4, starting from a modified BKP hierarchy, which can be viewed as the B\"acklund transformation of above extended BKP hierarchy, we construct integrable semi-discrete analogues of two-component reduced Ostrovsky equation and of the short wave limit of a two-component DP equation. We conclude our paper by some comments and further topics in section 5.
\section{The Lax pairs}
Eq. (\ref{2compVE2}) represents a conservation law, which can be used to define a hodograph (reciprocal) transformation
$(x,t)\rightarrow (y,s)$ by
\begin{equation}
dy=\rho dx-\rho udt,\quad ds=dt\,,
\end{equation}
then we have
\begin{equation}
\partial _{y}=\rho ^{-1}\partial_x,\quad \partial _{s}=\partial
_{t}+u\partial _{x}\,. \label{conversion1}
\end{equation}
By using above conversion formulas, we have the new conservative law
\begin{equation}
(\rho ^{-1})_{s}=u_{y}\,, \label{conservation2}
\end{equation}
or
\begin{equation}
\phi_{s}=u_{y}\,, \label{conservation3}
\end{equation}
by defining $\phi=\rho^{-1}$. Note that Eq. (\ref{2compVE1}) can be rewritten as
\begin{equation}
\rho u_{ys} -3u+ c (\rho-1)=0\,,
\end{equation}
which, in turn, to be
\begin{equation}
 \phi_{ss} -(3u+ c)\phi+ c=0\,, \label{SK}\,.
\end{equation}
As shown in \cite{Hone-Wang,Gordoa-Pickering}, Eqs. (\ref{conservation3}) and (\ref{SK}) belongs to the first negative flow in the Sawada-Kotera hierarchy. The corresponding Lax pair for $c=1$ is of third order, which can be expressed as
\begin{equation}
 \Psi _{sss}-\left( 3u+1\right) \Psi _{s} =\frac{1}{\lambda }\Psi\,, \label{Laxscaler2}
\end{equation}
\begin{equation}
 \Psi _{y}-\lambda (\phi\Psi _{ss}-\phi_{s}\Psi _{s}) =0\,.\label{Laxscaler1}
\end{equation}
The above Lax pair can be rewritten in a matrix form
\begin{equation}
\Psi _{y}=U\Psi ,\quad \Psi _{s}=V\Psi \,,  \label{Laxmatrix1}
\end{equation}%
with
\begin{equation}
\displaystyle U=\left(
\begin{array}{ccc}
0 & -\lambda \phi_{s} & \lambda \phi \\
\phi & \lambda & 0 \\
\phi_{s} & \phi & \lambda%
\end{array}%
\right)\,,
\end{equation}

\begin{equation}
\displaystyle V=\left(
\begin{array}{ccc}
0 & 1 & 0 \\
0 & 0 & 1 \\
\frac{1}{\lambda } & 3u+1 & 0%
\end{array}%
\right)\,.
\end{equation}

Applying the hodograph (reciprocal) transformation (\ref{conversion1}) to (\ref{Laxmatrix1}), we find
\begin{equation}
\Phi _{x}=U\Phi ,\quad \Phi _{t}=V\Phi \,, \label{Laxmatrix2}
\end{equation}%
with
\begin{equation}
\displaystyle U=\left(
\begin{array}{ccc}
0 & -\lambda u_{x} & \lambda \\
1 & \lambda \rho & 0 \\
u_{x} & 1 & \lambda \rho%
\end{array}%
\right)\,,
\end{equation}

\begin{equation}
\displaystyle V=\left(
\begin{array}{ccc}
0 & 1+\lambda uu_{x} & -\lambda u \\
-u & -\lambda \rho u & 1 \\
\frac{1}{\lambda }-uu_{x} & 2u+1 & -\lambda \rho u%
\end{array}%
\right)\,.
\end{equation}
It is easy to find that the zero-curvature condition for (\ref{Laxmatrix2}) yields the two-component reduced Ostrovsky equation (\ref{2compVE1})--(\ref{2compVE2}).

As the link of Lax pairs found by Hone and Wang in \cite{Hone-Wang} between the reduced Ostrovsky equation and the short wave limit of the DP equation, by considering the second component in above matrix form (\ref{Laxmatrix2}), we have
\begin{eqnarray}
&&\psi_{xxx}=2 \lambda \rho \psi_{xx}+\left(2 \lambda \rho_x
-\lambda^2 \rho^2\right) \psi_x+\left(\lambda m+\lambda \rho_{xx}-\lambda^2 \rho \rho_x\right) \psi,\\
&&\psi_t=\frac1{\lambda} \psi_{xx}-\left(u+\rho\right) \psi_x+\left(u_x-\rho_x\right)\psi\,.
\end{eqnarray}
The compatibility condition $\psi_{xxxt}=\psi_{txxx}$ gives the short wave limit of the two-component DP equation
(\ref{2compVE2})--(\ref{2compVE3}).
\section{Bilinear equation and $N$-soliton solution for the two-component reduced Ostrovsky equation}
\subsection{Bilinear equation}
The bilinear equation
\begin{equation}
[(D_{x_{-3}}-D_{x_{-1}}^3)D_{x_{1}}+3D_{x_{-1}}^2]\tau\cdot \tau=0\,,\label{BKP-bilinear}
\end{equation}
is a dual bilinear equation
\begin{equation}
[(D_{x_{3}}-D_{x_{1}}^3)D_{x_{-1}}+3D_{x_{1}}^2]\tau\cdot \tau=0\,,\label{BKP-bilinear2}
\end{equation}
which belongs to the extended BKP hierarchy \cite{HirotaBook,Jimbo-Miwa,Hirota-sww}.  It has been shown in \cite{FMO-VE} that this bilinear equation
yields the reduced Ostrovsky equation (\ref{vakhnenko}) through a hodograph transformation. Based on this finding, an integrable discretization of the reduced Ostrovsky equation (\ref{vakhnenko}) was constructed in \cite{FMO-VE-discrete}.

Impose a pseudo-3 reduction by requesting $D_{x_{-3}}  =c D_{x_{-1}}$ and assume $y=x_1$, $s=x_{-1}$, Eq. (\ref{BKP-bilinear}) is reduced to
\begin{equation} \label{2compVE-bilinear}
(D_{y}D_{s}^3-c D_{y}D_{s}-3D_{s}^2)\tau \cdot \tau=0\,.
\end{equation}
By using the relations
\begin{eqnarray*}
  \frac{D_{y}D_{s}^3 \tau \cdot \tau}{\tau^2} &=& 2 (\ln \tau)_{ysss} + 12 (\ln \tau)_{ss} (\ln \tau)_{ys}\,, \\
   \frac{D_{y}D_{s}\tau \cdot \tau}{\tau^2} &=& 2(\ln \tau)_{ys}\,, \quad
  \frac{D_{s}^2\tau \cdot \tau}{\tau^2} = 2(\ln \tau)_{ss}\,,
\end{eqnarray*}
Eq. (\ref{2compVE-bilinear}) is converted to
\begin{equation}
2(\ln \tau )_{ysss}=6(\ln \tau )_{ss}\left( 1-2(\ln \tau )_{ys}\right)
+2c(\ln \tau )_{ys},. \label{BKP-derivative}
\end{equation}
Introducing a dependent variable transformation
\begin{equation} \label{u-trf}
u=-2(\ln \tau )_{ss}\,,
\end{equation}
and a hodograph transformation
\begin{equation}  \label{hodo-trf}
 x=y-2(\ln \tau )_{s},\quad t=s\,,
\end{equation}
we then have
\begin{equation}
\frac{\partial x}{\partial y}=\rho ^{-1},\quad \frac{\partial x}{\partial s}=u
\end{equation}
by defining $\rho ^{-1}=1-2(\ln \tau )_{ys}$. Obviously, we have
\begin{equation}
\left(\rho^{-1} \right)_{s}=-2(\ln \tau )_{yss}=u_y\,,
\end{equation}
which, in turn, becomes
\begin{equation}
\rho_{s}=-\rho^2u_y=-\rho u_x\,.
\end{equation}
Furthermore, referring to the hodograph transformation and the resulting conversion formula (\ref{conversion1}), we obtain
\begin{equation}
\rho_t + u \rho_x =-\rho u_x\,,
\end{equation}
which is exactly Eq. (\ref{2compVE2}).
On the other hand, the dependent variable transformation (\ref{u-trf}) converts Eq. (\ref{BKP-derivative}) into
\begin{equation}
\rho u_{ys}=3u+c(1-\rho)\,.
\end{equation}
With the use of the conversion formula (\ref{conversion1}) by hodograph transformation, we have
\begin{equation}
(u _{t}+uu_{x})_x=3u+ c(1-\rho)\,,
\end{equation}
which is exactly Eq. (\ref{2compVE1}). In summary, the bilinear equation (\ref{2compVE-bilinear}) derives the two-component reduced Ostrovsky equation (\ref{2compVE1})--(\ref{2compVE2}) through the transformations (\ref{u-trf}) and (\ref{hodo-trf}).
\begin{remark}
A similar pseudo-3 reduction $D_{x_{3}}  =D_{x_{1}}$ acting on the bilinear equation (\ref{BKP-bilinear2}) leads to the shallow water waves
\cite{Hirota-Satsuma}
\begin{equation}
u _{t}-u_{txx}-3uu_{t}+3u_x \int_x^{\infty} u_t\,dx+ 3u_x=0\,,
\end{equation}
through variable transformations $x=x_1$, $t=x_{-1}$, $u=2(\ln \tau)_{xx}$.
\end{remark}
\begin{remark}
If $c=0$, the reduction becomes period 3 reduction satisfying $D_{x_{-3}} =0$, the resulting bilinear equation gives the reduced Ostrovsky equation
(\ref{vakhnenko}). Therefore, the reduced Ostrovsky equation can be viewed as a limiting case of the two-component reduced Ostrovsky equation as $c \to 0$. When $c=0$, even if the variable $\rho$ does not occur in the reduced Ostrovsky equation, it actually exists implicitly, which is embedded in the hodograph transformation.
\end{remark}
\subsection{$N$-soliton solution for the two-component reduced Ostrovsky equation (\ref{2compVE1})--(\ref{2compVE2})}
It is known that both the bilinear equations (\ref{BKP-bilinear})--(\ref{BKP-bilinear2}) admit a pfaffian-type solution \cite{HirotaBook,FMO-VE,FMO-VE-discrete}
\begin{equation}
\tau =\mathrm{Pf}(a_{1},a_{2},\cdots ,a_{2N}),
\end{equation}%
where the elements of pfaffian are defined by
\begin{equation}
\mathrm{Pf}(a_{i},a_{j})=c_{i,j}+\frac{p_{i}-p_{j}}{p_{i}+p_{j}}\varphi
_{i}\varphi _{j}\,,
\end{equation}
with
\begin{equation*}
c_{i,j}=-c_{j,i}\,, \quad \varphi _{i}=\exp (p_{i}^{-3}x_{-3}+p_{i}^{-1}x_{-1}+p_i x_1+p_{i}^{3}x_{3}+\xi_{i0}) \,.
\end{equation*}
Similar to the 3 reduction of the BKP hierarchy, to realize the pseudo-3 reduction $D_{x_{-3}}=cD_{x_{-1}}$, we need to impose a constraint
on the parameters of the general pfaffian solution, i.e.,
\begin{equation} \label{BKP-restriction}
c_{i,j}=\delta_{j,2N+1-i}c_i,
\quad c_{2N+1-i}=-c_i\,,
\end{equation}
and
\begin{equation}
p_i^{-3} +  p_{2N+1-i}^{-3} = c (p_i^{-1} +  p_{2N+1-i}^{-1})\,.
\end{equation}
Note that the pfaffian $\tau$ can be rewritten as
$$
\tau = \left(\prod_{i=1}^{2N}\varphi_i\right)
{\rm Pf}\pmatrix{\displaystyle
\frac{\delta_{j,2N+1-i}}{\varphi_i\varphi_{2N+1-i}}c_{i}
+\frac{p_i-p_j}{p_i+p_j}}\,,
$$
it can be easily shown that $\tau$ satisfies
\begin{equation}
 \partial_{x_{-3}}\tau=c\partial_{x_{-1}}\tau\,.
 \end{equation}
Under this reduction, the variable $x_{-3}$ becomes a dummy variable, which can be viewed as a constant. Summarizing the results in question, we can present the $N$-soliton solution by the following theorem.
\begin{theorem}
The two-component reduced Ostrovsky equation (\ref{2compVE1})--(\ref{2compVE2}) admits the following $N$-soliton solution in parametric form
\begin{equation}
u=-2(\ln \tau)_{ss}, \quad \rho=(1-2(\ln\tau)_{ys})^{-1}, \quad x=y-2(\ln \tau)_{s}\,, \quad t=s\,,
\end{equation}
where $\tau$ is a pfaffian
\begin{equation}
\tau = \mathrm{Pf} (a_1,a_2 \cdots, a_{2N})\,,
\end{equation}
whose elements are defined by
 \begin{equation}  \label{2comVE_pf1}
\mathrm{Pf}(a_i,a_j)= \delta_{j,2N+1-i}c_i+\frac{p_i-p_j}{p_i+p_j} e^{\xi_i+\xi_j}\,.
\end{equation}
Here $\xi_j=p_j y + p_j^{-1} s+\xi_{j0}$ with the wave numbers $p_j$ ($j=1,\cdots, 2N$)
satisfy a condition
\begin{equation}
p_i^{-3} +  p_{2N+1-i}^{-3} = c (p_i^{-1} +  p_{2N+1-i}^{-1})\,.
\end{equation}
\end{theorem}
\subsection{One- and two-soliton solutions}
In this subsection, we provide one- and two-soliton for the two-component reduced Ostrovsky equation (\ref{2compVE1})--(\ref{2compVE2}) and give a detailed analysis for their properties. \\
\noindent {\bf One-soliton} \\
For $N=1$, we have
\begin{equation}
 \tau={\rm Pf}(1,2)=c_1+\frac{p_1-p_2}{p_1+p_2}e^{\xi_1+\xi_{2}}\,.
\end{equation}
Let $c_1=1$, $\eta_1=$$\xi_1+\xi_{2} + \ln (p_1-p_2)-\ln (p_1+p_2)$, $p^{-1}_1+p^{-1}_2=k_1$, we then have
$p_1p_2=(k_1^2-c)/3$ since $p^{-3}_1+p^{-3}_2=c(p^{-1}_1+p^{-1}_2)$.  $\tau$ can be rewritten as
\begin{equation}
  \tau = 1+e^{\eta_1}
   = 1+ e^{k_1 s + \frac{3k_1}{k_1^2-c} y  + \eta_{10}}\,.
\end{equation}
Therefore, we have the parametric form of the one-soliton solution
\begin{equation}
u=-\frac{k_1^2}{2} {\rm sech}^2 \frac{\eta_1}{2}\,,
 \label{2comVE1solitona}
\end{equation}
\begin{equation}  \label{2comVE1solitonc}
\rho=\left(1- \frac{3k_1^2}{2(k_1^2-c)} {\rm sech}^2 \frac{\eta_1}{2}\right)^{-1}\,,
\end{equation}
\begin{equation}
x=y-\frac{2k_1e^{\eta_1}}{1+e^{\eta_1}}\,, \quad t=s\,.
 \label{2comVE1solitonb}
\end{equation}
Eq. (\ref{2comVE1solitona}) represents a soliton  of amplitude $k_1^2/2$ with velocity $-(k_1^2-c)/3$ for $u$-field.
The regularity of the solution depends on Eq. (\ref{2comVE1solitonc}).
Notice that $\rho \to 1$ as $y \to \pm \infty$, and it attains an extreme value of $-2(k_1^2-c)/(k_1^2+2c)$ at the peak point of the soliton when $\eta_{1}=0$,
it is not difficult to find that if $c>0$ and $k_1^2-c < 0$, or if $c<0$ and $k_1^2+2c <0$, the solution is regular.
Two examples for {\textbf{case (a)}: $k_1=1.0$, $c=2.0$ and {\textbf{case (b)}: $k_1=1.0$, $c=-2.0$ are illustrated in Fig. 1 and Fig. 2, respectively.
Even though the $u$-field has the same amplitude for both cases, the $\rho$-field is quite different. The amplitude of $\rho$ is smaller that the asymptotic value of $1$ at $\pm \infty$ for case (a), while it is larger than $1$ for case (b). Moreover, the soliton moves to the right with velocity $1/3$ for case (a) and to the left with velocity $-1$ for case (b).
\begin{figure}[htbp]
\centerline{
\includegraphics[scale=0.35]{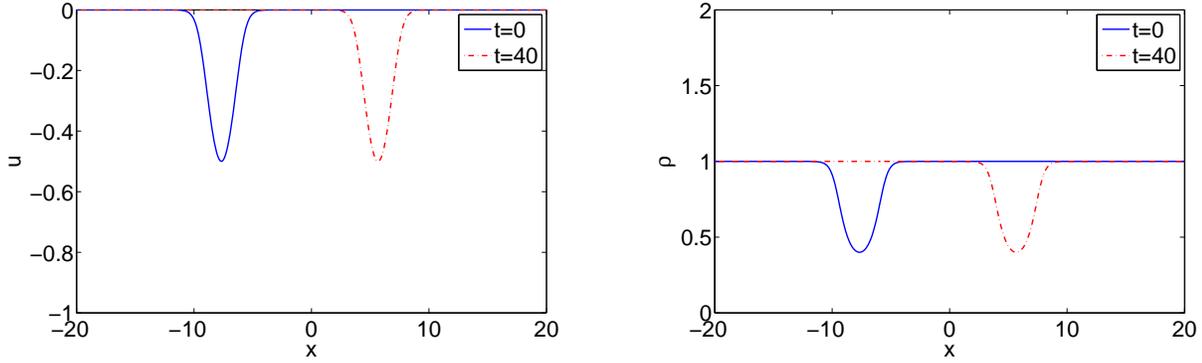}\quad
\includegraphics[scale=0.35]{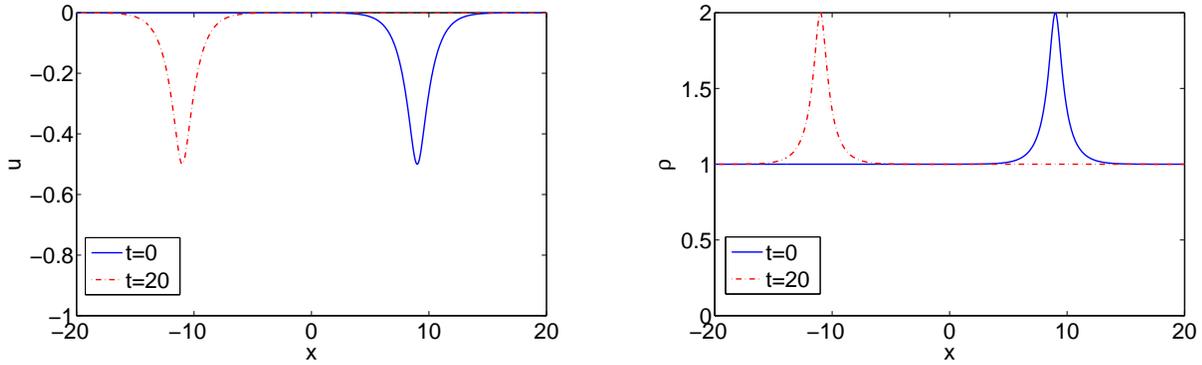}}
\caption{A smooth soliton for two-component reduced Ostrovsky equation with $k_1=1.0$, $c=2.0$: (a) profile of $u$, (b) profile of $\rho$.}
\label{fig1}
\end{figure}

\begin{figure}[htbp]
\centerline{
\includegraphics[scale=0.35]{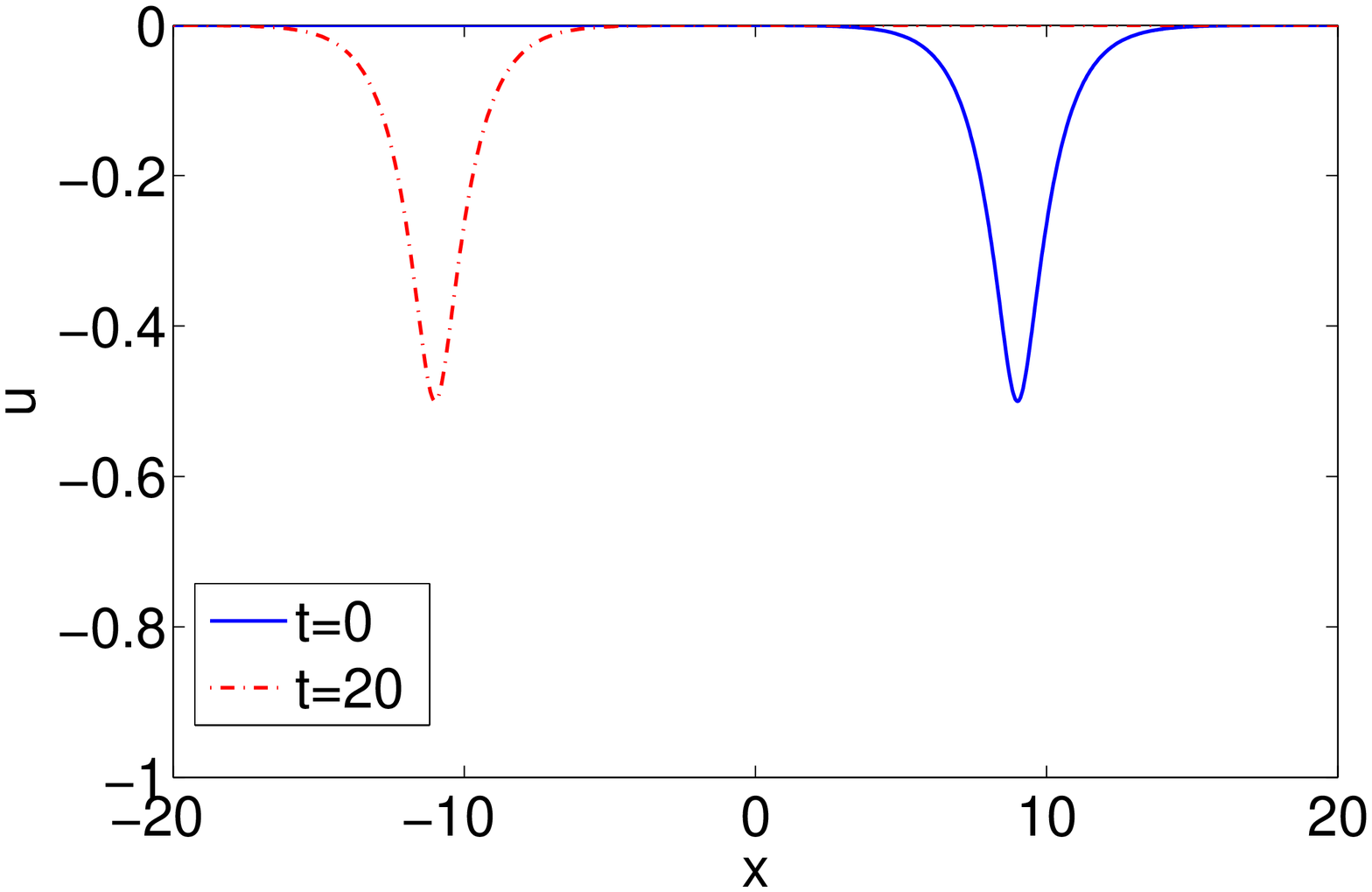}\quad
\includegraphics[scale=0.35]{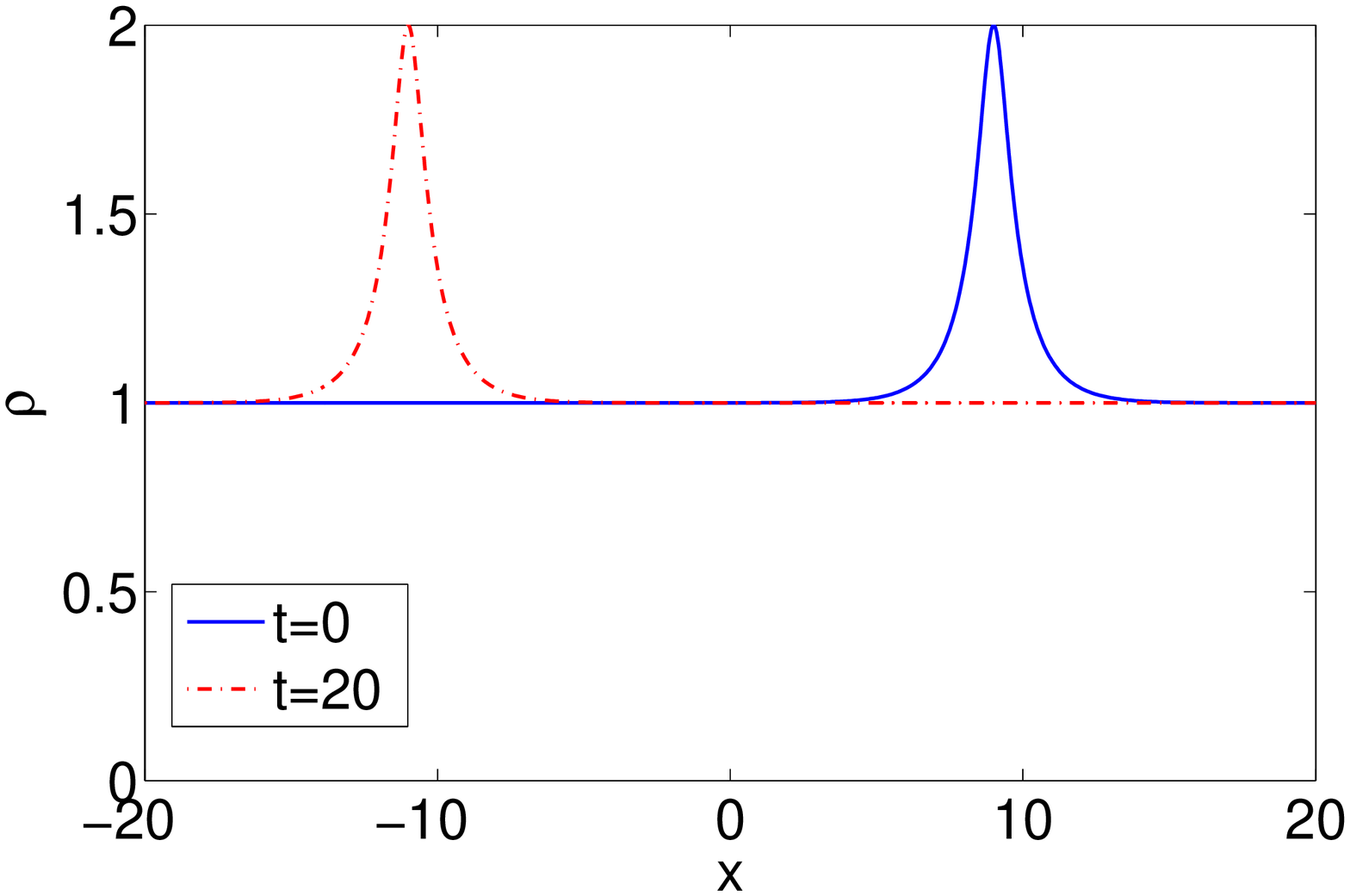}}
\caption{A smoth soliton for two-component reduced Ostrovsky equation with $k_1=1.0$, $c=-2.0$: (a) profile of $u$, (b) profile of $\rho$.}
\label{fig2}
\end{figure}

\begin{remark}
When $c=0$, the two-component reduced Ostrovsky equation  becomes simply the reduced Ostrovsky equation, and the one-soliton solution is always of loop type since $\rho^{-1}$ has alwasy two zeros. Whereas, the two-component reduced Ostrovsky equation has the regular solution depending on the values of $c$ and wave number $\kappa_1$. \end{remark}
\begin{remark}
In compared with the reduced Ostrovsky equation which only admits the left-moving soliton
solution, the two-component reduced Ostrovsky equation may have both the left-moving and right-moving soliton solutions.
To be more specific, if $k_1^2 -c >0$, it has left-moving soliton, whereas, if $k_1^2 -c <0$, it has right-moving soliton. However, the soliton solution does not exist when $k_1^2 -c =0$.
\end{remark}
\noindent {\bf Two-soliton} \\
By choosing $c_1=c_2=1$, we have the tau function for two-soliton solution ($N=2$)
\begin{eqnarray*}
\fl \tau &=&{\rm Pf}(1,2,3,4)
={\rm Pf}(1,2){\rm Pf}(3,4)-{\rm Pf}(1,3){\rm Pf}(2,4)
+{\rm Pf}(1,4){\rm Pf}(2,3)\\
\fl &=&\frac{p_1-p_2}{p_1+p_2}e^{\xi_1+\xi_{2}} \times
 \frac{p_3-p_4}{p_3+p_4}e^{\xi_3+\xi_{4}}
-\frac{p_1-p_3}{p_1+p_3}e^{\xi_1+\xi_{3}} \times
 \frac{p_2-p_4}{p_2+p_4}e^{\xi_2+\xi_{4}}\\
\fl &&\qquad +\left(1+\frac{p_1-p_4}{p_1+p_4}e^{\xi_1+\xi_4}\right)
\left(1+\frac{p_2-p_3}{p_2+p_3}e^{\xi_2+\xi_3}\right)\,,
\end{eqnarray*}
under the condition
\begin{equation}
p_1^{-3}+p_{4}^{-3}
=c(p_1^{-1}+p_{4}^{-1})\,, \quad
p_2^{-3}+p_{3}^{-3}
=c(p_2^{-1}+p_{3}^{-1})\,.
\end{equation}
Similarly, the above $\tau$-function
can be rewritten as
\begin{equation}
 \tau=1+e^{\eta_1}
+e^{\eta_2}+b_{12}e^{\eta_1+\eta_2}\,,
\end{equation}
with
\begin{equation}
b_{12}=
\frac{(p_1-p_2)(p_1-p_3)(p_4-p_2)(p_4-p_3)}{(p_1+p_2)(p_1+p_3)(p_4+p_2)(p_4+p_3)}
\,,
\end{equation}
by having $\eta_1=$$\xi_1+\xi_{3} + \ln (p_1-p_3)-\ln (p_1+p_3)$, $\eta_2=$$\xi_2+\xi_{4} + \ln (p_2-p_4)-\ln (p_2+p_4)$. Furthermore, if we let $p^{-1}_1+p^{-1}_4=k_1$, $p^{-1}_2+p^{-1}_3=k_2$, we then have
\begin{equation}
\eta_i= k_i s + \frac{3k_i}{k_i^2-c} y  + \eta_{i0}
\,,
\end{equation}
for $i=1,2$ and
\begin{equation}
b_{12}=
\frac{(k_1-k_2)^2(k^2_1-k_1k_2+k_2^2-3c)}{(k_1+k_2)^2(k^2_1+k_1k_2+k_2^2-3c)}
\,.
\end{equation}
To avoid the singularity of the soliton solution, the condition $k_i^2+2c < 0$ ($i=1,2$)
need to be satisfied. In regard to the interactions of two solitons, there are either catch-up collision or head-on collision
depending on the values of parameters discussed previously.  Furthermore, the
collision is always elastic, there is no change in shape and amplitude of
solitons except a phase shift. In Fig. 3, we illustrate
the contour plot for the collision of two solitons, and in Fig. 4, the
profiles before and after the collision. The parameters are taken as $%
c=-2.0$, $k_1=1.0$ and $k_2=1.6$.

\begin{figure}[htbp]
\centerline{
\includegraphics[scale=0.35]{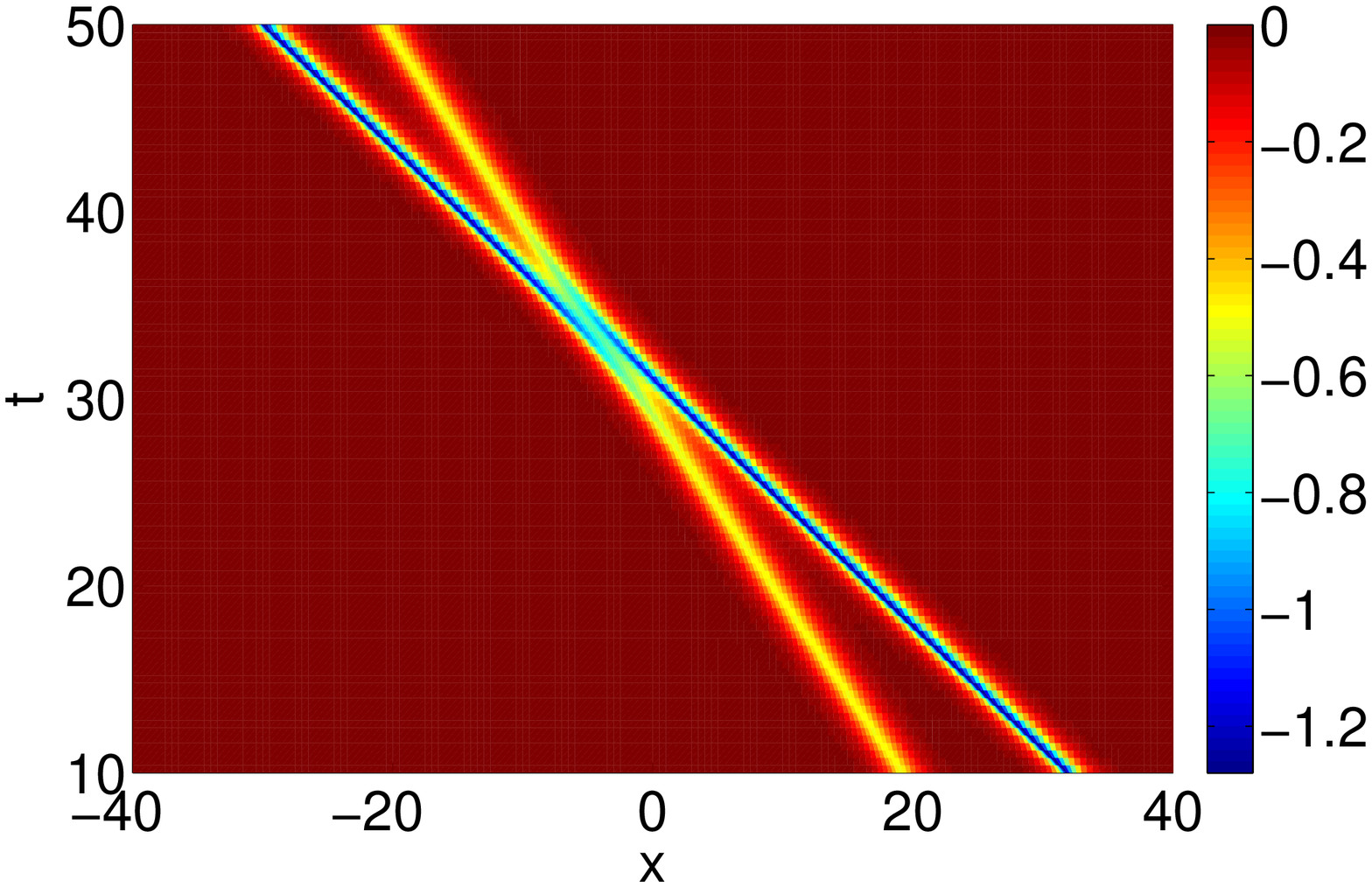}\quad
\includegraphics[scale=0.35]{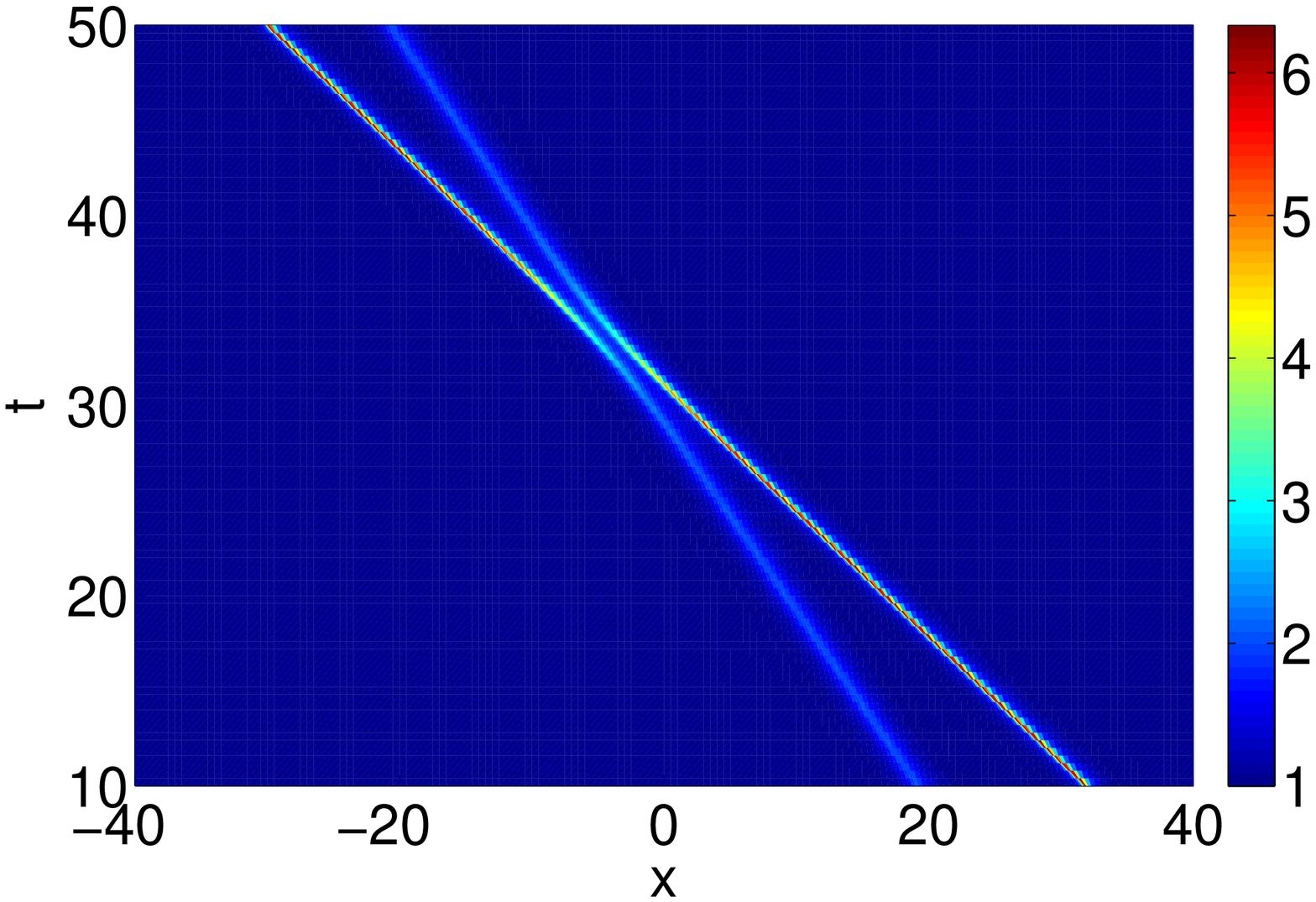}}
\caption{Collision between two solitons for two-component reduced Ostrovsky equation with $k_1=1.0$, $k_2=1.6$ $c=-2.0$: (a) contour plot of $u$, (b)contour plot of $\rho$}
\label{fig3}
\end{figure}

\begin{figure}[htbp]
\centerline{
\includegraphics[scale=0.35]{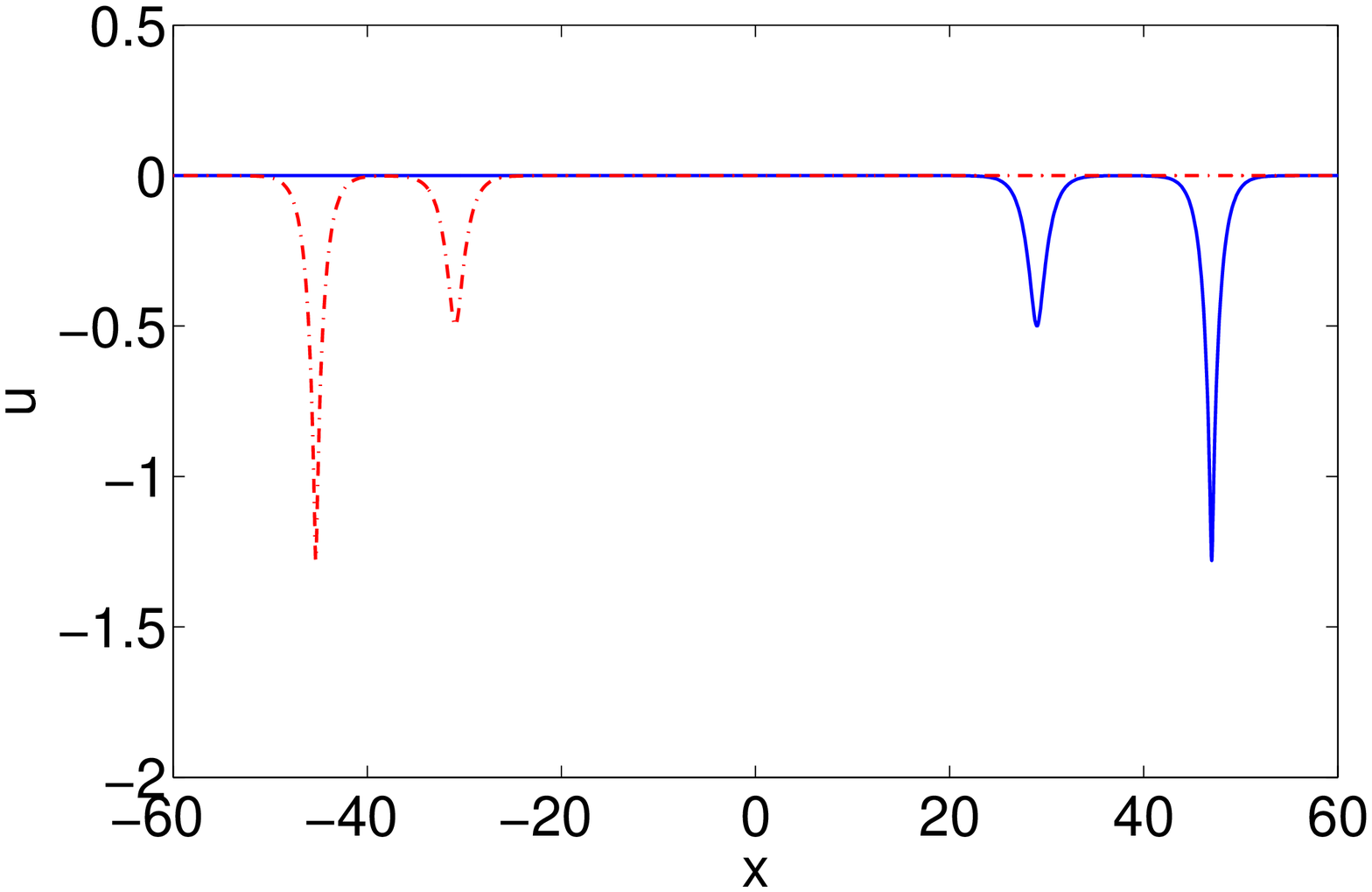}\quad
\includegraphics[scale=0.35]{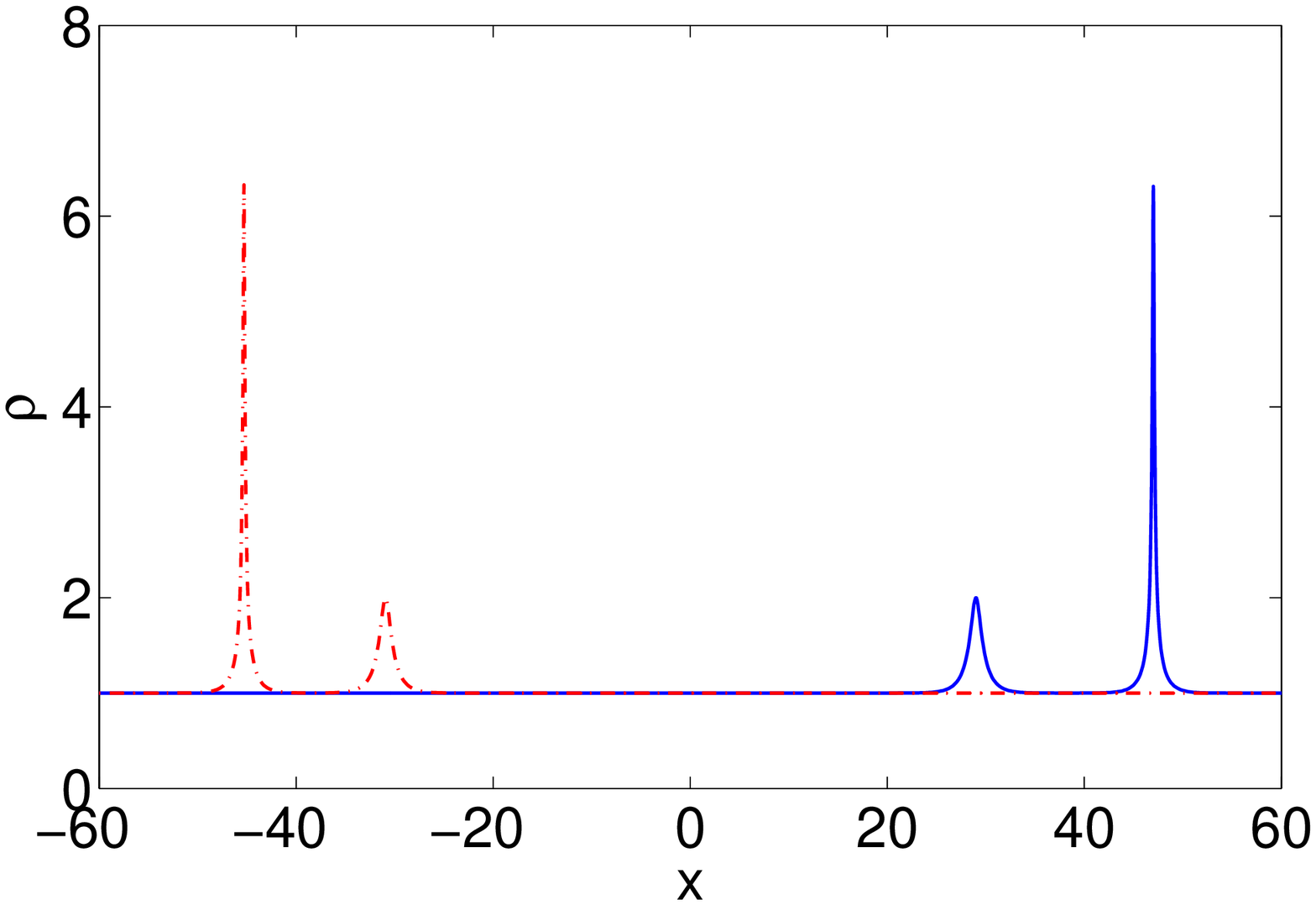}}
\caption{Collision between two solitons for two-component reduced Ostrovsky equation with $k_1=1.0$, $k_2=1.6$ $c=-2.0$: (a) profile of $u$, (b) profile of $\rho$.}
\label{fig4}
\end{figure}


\section{Integrable semi-discretization of the two-component reduced
Ostrovsky equation}
We could construct a semi-discrete analogue of the two-component reduced
Ostrovsky equation based on the B\"aclund transformation of the extended BKP hierarchy. For the sake of simplicity, here we take $c=1$ without loss of generality. The starting point is a bilinear equation associated with the modified
BKP hierarchy
\begin{equation}
\left( (D_{s}-b)^{3}-(D_{r}-b^{3})\right) \tau _{l+1}\cdot \tau _{l}=0\,.
\label{bilinear_BKP2}
\end{equation}
This bilinear equation can be viewed as a B\"aclund transformation of the extended BKP hierarchy. It admits a pfaffian type solution of the form $\tau _{l}=\mathrm{Pf}%
(1,2,\cdots ,2N)_{l}$ whose elements are determined by
\begin{equation}
(i,j)_{l}=c_{i,j}+\frac{p_{i}-p_{j}}{p_{i}+p_{j}}\varphi
_{i}^{(0)}(l)\varphi _{j}^{(0)}(l)\,,  \label{bilinear_BKP1}
\end{equation}%
where $c_{i,j}=-c_{j,i}$ and
\[
\varphi _{i}^{(n)}(l)=p_{i}^{n}\left( \frac{1+bp_{i}}{1-bp_{i}}\right)
^{l}e^{\xi _{i}},\quad \xi _{i}=p_{i}^{-1}s+p_{i}^{-3}r+\xi _{i0}\,.
\]
Note that if we take $c_{i,j}$ as in Eq.(\ref{BKP-restriction}), $\tau_l$ is rewritten
\[
\tau _{l}=\left( \prod_{i=1}^{2N}\varphi _{i}^{(0)}(l)\right) \mathrm{Pf}%
\pmatrix{\displaystyle
\frac{\delta_{j,2N+1-i}}{\varphi_i^{(0)}(l)\varphi_{2N+1-i}^{(0)}(l)}
c_{i} +\frac{p_i-p_j}{p_i+p_j}}\,,
\]
so by imposing a reduction condition
\[
\frac{1}{p_{i}^{3}}+\frac{1}{p_{2N+1-i}^{3}}=\frac{1}{p_{i}}+\frac{1}{%
p_{2N+1-i}}\,,
\]%
we can easily show that the pfaffian $\tau _{l}$ satisfies
\begin{equation}
\partial _{r}\tau _{l}=\partial _{s}\tau _{l}\,.
\end{equation}%
Therefore Eq. (\ref{bilinear_BKP2}) is reduced into
\begin{equation}
(D_{s}^{3}-3bD_{s}^{2}+(3b^{2}-1)D_{s})\tau _{l+1}\cdot \tau _{l}=0\,,
\label{Bilinear_semiVE}
\end{equation}
based on which we will derive the integrable semi-discretization.
First, we introduce a discrete hodograph transformation
\begin{equation}
x_{l}=2lb-2(\ln \tau _{l})_{s},\quad t=s\,,  \label{sd_hodograph_trf}
\end{equation}
and a dependent variable transformation
\begin{equation}
u_{l}=-2(\ln \tau _{l})_{ss},  \label{sd_u_trf}
\end{equation}
\begin{equation}
\rho _{l}=\left(1-b^{-1}\left(\ln \frac{\tau _{l+1}}{\tau _{l}}\right)_{ss}\right)^{-1},
\label{sd_rho_trf}
\end{equation}
it then follows that the nonuniform mesh, which is defined by $\delta
_{l}=x_{l+1}-x_{l}$, can be expressed as
\begin{equation}
\delta _{l}=2b-2\left( \ln \frac{\tau _{l+1}}{\tau _{l}}\right) _{s},
\label{mesh}
\end{equation}
which is related to $\rho _{l}$ by
\begin{equation}
\rho _{l}=\frac{2b}{\delta _{l}}.
\end{equation}
 Differentiating Eq. (\ref{mesh}) with respect to $s$, one obtains
\begin{equation}
\frac{d\delta _{l}}{ds}=-2\left( \ln \frac{\tau _{l+1}}{\tau _{l}}\right)
_{ss}=u_{l+1}-u_{l}\,.  \label{sd1_VE2}
\end{equation}
which is equivalent to
\begin{equation}
\frac{d\rho _{l}^{-1}}{ds}=\frac{u_{l+1}-u_{l}}{2b}\,.
\end{equation}
Dividing $\tau _{l+1}\tau _{l}$ on both sides of Eq.(\ref{Bilinear_semiVE}) and using the following relations
\[
\frac{D_{s}\tau _{l+1}\cdot \tau _{l}}{\tau _{l+1}\tau _{l}}=\left( \ln
\frac{\tau _{l+1}}{\tau _{l}}\right) _{s}\,,
\]

\[
\frac{D^2_{s} \tau_{l+1} \cdot \tau_{l}}{\tau_{l+1}\tau_{l}} = \left(\ln
(\tau_{l+1}\tau_{l})\right)_{ss} +\left(\left(\ln \frac{\tau_{l+1}}{\tau_{l}}%
\right)_s\right)^2 \,,
\]
\[
\frac{D_{s}^{3}\tau _{l+1}\cdot \tau _{l}}{\tau _{l+1}\tau _{l}}=\left( \ln
\frac{\tau _{l+1}}{\tau _{l}}\right) _{sss}+3\left( \ln \frac{\tau _{l+1}}{%
\tau _{l}}\right) _{s}\left( \ln (\tau _{l+1}\tau _{l})\right) _{ss}+\left(
\left( \ln \frac{\tau _{l+1}}{\tau _{l}}\right) _{s}\right) ^{3}\,,
\]%
one obtains 
\begin{eqnarray}
\left( \ln \frac{\tau _{l+1}}{\tau _{l}}\right) _{sss} &=&(1-b^{2})\left(
\ln \frac{\tau _{l+1}}{\tau _{l}}\right) _{s}+\left( b-\left( \ln \frac{\tau
_{l+1}}{\tau _{l}}\right) _{s}\right)  \nonumber \\
&&\left[ 3\left( \ln (\tau _{l+1}\tau _{l})\right) _{ss}-\left( \ln \frac{%
\tau _{l+1}}{\tau _{l}}\right) _{s}\left( 2b-\left( \ln \frac{\tau _{l+1}}{%
\tau _{l}}\right) _{s}\right) \right] \,,  \label{BL_semiVE2}
\end{eqnarray}%
which is converted into
\begin{equation}
\frac{d}{ds}(u_{l+1}-u_{l})=\frac{3}{2}\delta _{l}(u_{l}+u_{l+1})-\frac{1}{4}%
\delta _{l}(\delta _{l}^{2}-4)+2b^{3}-2b\,,
\label{sd2-VE1}
\end{equation}
by Eqs. (\ref{sd_u_trf}) and (\ref{mesh}). In summary we have the following theorem
\begin{theorem}
The bilinear equation
\[
(D_{s}^{3}-3bD_{s}^{2}+(3b^{2}-1)D_{s})\tau _{l+1}\cdot \tau _{l}=0\,
\]%
determines a semi-discrete analogue of the two-component reduced Ostrovsky
equation (\ref{2compVE1})--(\ref{2compVE2})
\begin{equation} \label{sd-2comVE1}
\frac{d}{ds}(u_{l+1}-u_{l})=\frac{3}{2}\delta
_{l}(u_{l}+u_{l+1})-\frac{1}{4}\delta _{l}(\delta _{l}^{2}-4)+2b^{3}-2b\,,
\end{equation}
\begin{equation} \label{sd-2comVE2}
\frac{d\rho _{l}^{-1}}{ds}=\frac{u_{l+1}-u_{l}}{2b}\,
\end{equation}
by dependent variable transformations
\begin{equation}
u_{l}=-2(\ln \tau _{l})_{ss},  \quad \rho _{l}=\left(1-b^{-1}\left(\ln \frac{\tau _{l+1}}{\tau _{l}}\right)_{ss}\right)^{-1}\,,
\end{equation}
and a discrete hodograph transformation
\begin{equation}
x_{l}=2lb-2(\ln \tau _{l})_{s},\quad t=s\,.
\end{equation}
\end{theorem}
The nonuniform mesh, which is defined by $\delta_{l}=x_{l+1}-x_{l}$, is related to $\rho_l$ by
$\rho _{l}{\delta _{l}}={2b}$. Next, we show the continuous limit of semi-discrete two-component reduced Ostrovsky equation
(\ref{sd-2comVE1})--(\ref{sd-2comVE2}). Since
\[
\frac{\partial x}{\partial s}=\frac{\partial x_{0}}{\partial s}%
+\sum_{j=0}^{l-1}\frac{\partial \delta _{j}}{\partial s}=\frac{\partial x_{0}%
}{\partial s}+\sum_{j=0}^{l-1}(u_{j+1}-u_{j})\rightarrow u\,,
\]%
we then have
\[
\partial _{s}=\partial _{t}+\frac{\partial x}{\partial s}\partial
_{x}\rightarrow \partial _{t}+u\partial _{x}\,.
\]%
Then Eq. (\ref{sd-2comVE2}) is converted into
\[
\partial _{x}(\partial _{t}+u\partial _{x})\frac{1}{\rho}=u_y,
\]
which, in turn, becomes Eq. (\ref{2compVE2}).
By dividing $\delta _{l}$ on both sides of Eq. (\ref{sd2-VE1}), we have
\begin{equation}
\frac{1}{\delta _{l}}\frac{d}{ds}(u_{l+1}-u_{l})=\frac{3}{2}(u_{l}+u_{l+1})-%
\frac{1}{4}\delta _{l}^{2}+1-(1-b^{2})\rho _{l}\,.  \label{sd2-VE3}
\end{equation}%
Obviously, in the continuous limit, $b\rightarrow 0$ ($\delta
_{l}\rightarrow 0$), it converges to
\[
(\partial _{t}+u\partial _{x})u=3u\,+1-\rho\,,
\]
which is exactly Eq. (\ref{2compVE1}) with $c=1$.
It is interesting to note that we have
\begin{eqnarray}
&&\frac{1}{\delta _{l}}\frac{d}{ds}(u_{l+1}-u_{l})-\frac{1}{\delta _{l-1}}%
\frac{d}{ds}(u_{l}-u_{l-1})  \nonumber \\
&& \quad =\frac{3}{2}(u_{l+1}-u_{l-1})-\frac{1}{4}\left(
\delta _{l}^{2}-\delta _{l-1}^{2}\right)-(1-b^{2})(\rho _{l}-\rho _{l-1})\,,
\label{sd2-VE4}
\end{eqnarray}%
by taking a backward difference of Eq. (\ref{sd2-VE3}). Furthermore, by defining
\begin{equation*}
m_{l}=1-\frac{2}{\delta _{l}+\delta _{l-1}}\left( \frac{u_{l+1}-u_{l}}{%
\delta _{l}}-\frac{u_{l}-u_{l-1}}{\delta _{l-1}}\right)\,,
\end{equation*}%
by defining a forward difference operator and an average operator
\begin{equation*}
\Delta f_{l}=\frac{f_{l+1}-f_{l}}{\delta _{l}},\quad Mu_{l}=\frac{%
f_{l}+f_{l-1}}{2}\,,
\end{equation*}%
we can claim an integrable semi-discrete analogue of Eqs. (\ref{sd-2comVE1})--(\ref{sd-2comVE2}) as follows
\begin{theorem}
A semi-discrete analogue for the short wave limit of a two-component DP equation (\ref{2compVE2})--(\ref{2compVE3}) is of the form
\begin{equation} \label{sd-sw2DP1}
 \frac{d\,m_{l}}{d\,s}=m_{l}\left( -2M\Delta u_{l}-\frac{%
M(\delta _{l}\Delta u_{l})}{M\delta _{l}}+\frac{1}{2}(\delta _{l}-\delta
_{l-1})\right)+(1-b^{2})\frac{\rho _{l}-\rho _{l-1}}{M\delta _{l}}\,,
\end{equation}
\begin{equation} \label{sd-sw2DP3}
\frac{d\rho _{l}^{-1}}{ds}=\frac{u_{l+1}-u_{l}}{2b}\,,
\end{equation}
\begin{equation}
m_{l}=1-\frac{2}{\delta _{l}+\delta _{l-1}}\left( \frac{u_{l+1}-u_{l}}{%
\delta _{l}}-\frac{u_{l}-u_{l-1}}{\delta _{l-1}}\right)\,.  \label{sd-sw2DP2}
\end{equation}%
\end{theorem}
Its $N$-soliton solution is the same as the one of the two-component reduced Ostrovsky equation. In the continuous limit, $b\rightarrow 0$ ($\delta _{l}\rightarrow 0$), we
have
\[
2M\Delta u_{l} \to 2u_{x}\,,\quad \frac{M(\delta _{l}\Delta u_{l})}{%
M\delta _{l}} \to u_{x}\,,\quad \frac{\rho _{l}-\rho _{l-1}}{M\delta _{l}}%
\to \rho _{x}\,,
\]
then Eq. (\ref{sd-sw2DP2}) converges to
\[
m_{l}\rightarrow m=1-u_{xx}\,,
\]
while Eq. (\ref{sd-sw2DP1}) and  (\ref{sd-sw2DP3}) converge to
\begin{equation*}
(\partial _{t}+u\partial _{x})m=-3mu_{x}-\rho _{x}\,,
\end{equation*}%
and \begin{equation*}
(\partial _{t}+u\partial _{x}) \rho=-\rho u_{x}\,,
\end{equation*}%
which are exactly the short wave limit of a
two-component DP equation (\ref{2compVE2})--(\ref{2compVE3}).
\section{Conclusion and further topics}
In the present paper, we proposed a two-component generalization of the reduced Ostrovsky equation and its differential form, which can be viewed a short wave limit of a two-component DP equation. The integrability for both equations is assured by finding their Lax pairs. Moreover, we have shown that the proposed two-component reduced Ostrovsky equation can be reduced from an extended BKP hierarchy through a hodograph transformation under a pseudo 3-reduction. Based on this fact, its bilinear equation, as well as its $N$-soliton solution, is found. One- and two-soliton solutions are analyzed in details. We should emphasize that, in compared with the reduced Ostrovsky equation which only admits multi-valued (loop) soliton solution, the two-component reduced Ostrovsky equation, as well as its differential form, can have regular solutions depending on the spatial wave number and the value of $c$.

The integrable semi-discrete analogues for the two-component generalization of the reduced Ostrovsky equation and its differential form are constructed based on a B\"acklund transform of the extended BKP hierarchy by defining a discrete hodograph transform and mimicking pseudo 3-reduction in continuous case. The $N$-soliton solutions are also provided in terms of pfaffians. It would be interesting to apply integrable semi-discretizations as integrable self-adaptive moving mesh methods \cite{dCH,dCHcom,SPE_discrete1}  for numerical simulations of the two-component reduced Ostrovsky equation.

A two-component Camassa-Holm (2-CH) equation \cite{YoujinLMP,AGZJPA,ConstantinIvanov} and its short wave limit, also called two-component Hunter-Saxton (2-HS) equation \cite{Wunsch09,Lenells09,MoonLiu2012,LouFeng}, have been known for while and has drawn some attentions in mathematical physics. Both equations can be expressed by the same form
\begin{equation}
m_t + u m_x+2m u_x-\sigma \rho \rho_x=0,  \label{2CHa}
\end{equation}
\begin{equation}
\rho_t+(\rho u)_x=0,  \label{2CHb}
\end{equation}
except for the 2-CH equation $m= \kappa+u-u_{xx}$ and for the 2-HS equation $m= \kappa-u_{xx}$. A similar two-component DP equation has been proposed in \cite{2DPPop} but it seems not integrable. Does an integrable two-component DP equation share the same form as Eqs. (\ref{2compVE2})--(\ref{2compVE2}) except $m= 1+u-u_{xx}$. If this is true, then what is the Lax pair? We expect that the answers to these questions can be made clear in the near future.
\section*{Acknowledgment}
BF appreciates the comments and discussions with Professor Youjin Zhang and Professor Qingping Liu. The work of KM is partially supported by CREST, JST. The work of YO is partially supported by JSPS Grant-in-Aid for Scientific Research (B-24340029, C-15K04909) and for Challenging Exploratory Research (26610029).

\end{document}